%% file: spreadPaper.tex
\documentclass[11pt]{article}
\pdfoutput=1
\usepackage{fullpage}

\usepackage{url}
\usepackage{algorithm2e}
\usepackage{enumerate,color,graphicx}
\usepackage{array}
\usepackage[fleqn]{amsmath}
\usepackage{tabularx,booktabs,colortbl}

\begin{document}

\title{A Hybrid Model for Disease Spread and an Application to the
  SARS Pandemic}
\author{Teruhiko Yoneyama, Sanmay Das, and Mukkai Krishnamoorthy\\
 Rensselaer Polytechnic Institute\\
  Troy, NY 12180, USA}
\date{}

\maketitle
\begin{abstract}
  Pandemics can cause immense disruption and damage to communities and
  societies. Thus far, modeling of pandemics has focused on either
  large-scale difference equation models like the SIR and the SEIR
  models, or detailed micro-level simulations, which are harder to
  apply at a global scale. This paper introduces a hybrid model for
  pandemics considering both global and local spread of infections. We
  hypothesize that the spread of an infectious disease between regions
  is significantly influenced by global traffic patterns and the
  spread within a region is influenced by local conditions. Thus we
  model the spread of pandemics considering the connections between
  regions for the global spread of infection and population density
  based on the SEIR model for the local spread of infection. We
  validate our hybrid model by carrying out a simulation study for the
  spread of SARS pandemic of 2002-2003 using available data on
  population, population density, and traffic networks between
  different regions. While it is well-known that international
  relationships and global traffic patterns significantly influence
  the spread of pandemics, our results show that integrating these
  factors into relatively simple models can greatly improve the
  results of modeling disease spread.
\end{abstract}

\section{Introduction}
\label{section:intro}
\input{intro.tex}

\section{Modeling the Spread of Disease}
\label{section:model}
\input{model.tex}

\section{Calibration With Data}
\label{section:data}
\input{data.tex}

\section{Results}
\label{section:results}
\input{results.tex}

\section{Discussion}
\label{section:discussion}
\input{discussion.tex}



\input{spreadPaper.bbl}
\end{document}

%% file: intro.tex
Modeling of the spread of infectious disease typically falls into one of two categories. Analytically tractable models like the SEIR model are capable of capturing some globally important phenomena like the rate of spread of diseases using few parameters. However, they have a hard time reflecting differences in global spread due to local conditions. For example, it can be difficult to model different rates of spread in countries with different population densities and public health policies of variable strength and coordination. Network- or agent-based models are capable of reflecting details of individual conditions. However, modeling large-scale global disease-spread using such models often runs into methodological problems like overfitting because of the vast number of possible parameters. 

This paper proposes a granular, network-based hybrid model of disease spread in which individual regions are modeled as nodes in the network, and the spread of disease within nodes is modeled analytically (using a simplified derivative of the SEIR model) with the help of demographic parameters like population density. The properties of the network as a whole, like connectivity, are determined using real data on traffic between regions. We demonstrate the power of this approach by simulating the spread of SARS . One of the key takeaways is that the level of granularity has a significant effect on the success of network- or agent-based simulation models. For example, we show that modeling China as an individual node is unsuccessful, whereas breaking it up into constituent regions gives an impressive match to real infection data on SARS.

One of the great advantages of our model is its parsimony: it contains relatively few tweakable parameters compared with general agent-based models. At the same time it is capable of reproducing the important broad flows of disease. However, it is important to remember that exact reproduction of historical data is not the end-goal. Exceptions that do not correspond to real data provide insight into specific local phenomena that influenced the progression of a pandemic, such as an actual timing of the first infected case in a country.

\subsection{Related Work}

There is a vast literature on understanding the spread of disease using analytical and simulation models. In the next section we give a brief overview of the most common modeling methodologies, including differential equation models and simulation models, but here we discuss related research more generally. The most closely related to this work can be grouped into two categories. First, several researchers have simulated and analyzed the \emph{local} spread of SARS in 2002-2003 \cite{Huang2004,Li2004,Zhang2005}. In particular, Huang, \emph{et al} reproduce the situations in Singapore, Taipei, and Toronto individually, and compare with the actual transitions \cite{Huang2004}.  This also ties in to a significant existing literature on local modeling of historical pandemics, like the Influenza during the First World War (e.g. \cite{Chowell2006,Massad2006}). Other examples also abound: Jenvald, \emph{et al} use a virtual city based on Link\"{o}ping, Sweden, considering the number of schools, age distribution, and household type \cite{Jenvald2007}; Longini, \emph{et al} model population and contact processes based on Thailand census data, demographic information, and social network data \cite{Longini2005}; Kelso, \emph{et al} model a real community in the south west of Western Australia \cite{Kelso2009,Milne2008}.

The second category involves simulating global infection spread using international traffic data.  For example, several papers use air travel data to estimate connectivity in a network \cite{Colizza2007,Cooper2006,Ferguson2006,Flahault2006}. However, these authors typically simulate a hypothetical global pandemic, with a focus on intervention policies; the focus of our research is to validate the simulation with real historical data.

Much existing research simulates infection in networks with reasonable properties, but not necessarily based on existing real-world data.  For example, Bailey simulates epidemics in two dimensions, such as square grids \cite{Bailey1965}. Patel, \emph{et al} \cite{Patel2004} and Weycker, \emph{et al} \cite{Weycker2004} consider hypothetical populations of 10,000 persons, comprised of five communities of equal size, containing schools and neighborhoods. Vespignani, Pastor-Satorras, and co-authors simulate spreading infectious diseases with complex networks\cite{Boguna2003,Moreno2002,PastorSatorras20011,PastorSatorras20012,PastorSatorras20021,PastorSatorras20022}. Carrat, \emph{et al} \cite{Carrat2006}, Glass, \emph{et al} \cite{Glass2006}, and Eubank \cite{Eubank2002} also use generated complex networks for simulation.

Another major theme of research has been on the effects of prevention and/or mitigation strategies. These typically compare a ``base'' simulation and an alternative simulation which considers some proposed strategy. For example, Longini, \emph{et al} use stochastic epidemic simulations to investigate the effectiveness of targeted antiviral prophylaxis to contain influenza \cite{Longini2004}. Kelso, \emph{et al} simulate the effect of social isolation, such as school closure, individual isolation, workplace nonattendance, and reduction of contact \cite{Kelso2009}. Carrat, \emph{et al} explore the impact of interventions, such as vaccination, treatment, quarantine, and closure of schools and workplaces \cite{Carrat2006}. Germann, \emph{et al} simulate and compared the baseline and several combinations of mitigation methods \cite{Germann2006}. Patel, \emph{et al} use genetic algorithms to find optimal vaccination strategies \cite{Patel2004}. Weycker, \emph{et al} estimate the population-wide benefits of routine vaccination of children \cite{Weycker2004}.

\subsection{The SARS Pandemic}

\begin{figure}
\centering
\includegraphics[angle=90,width=\textwidth]{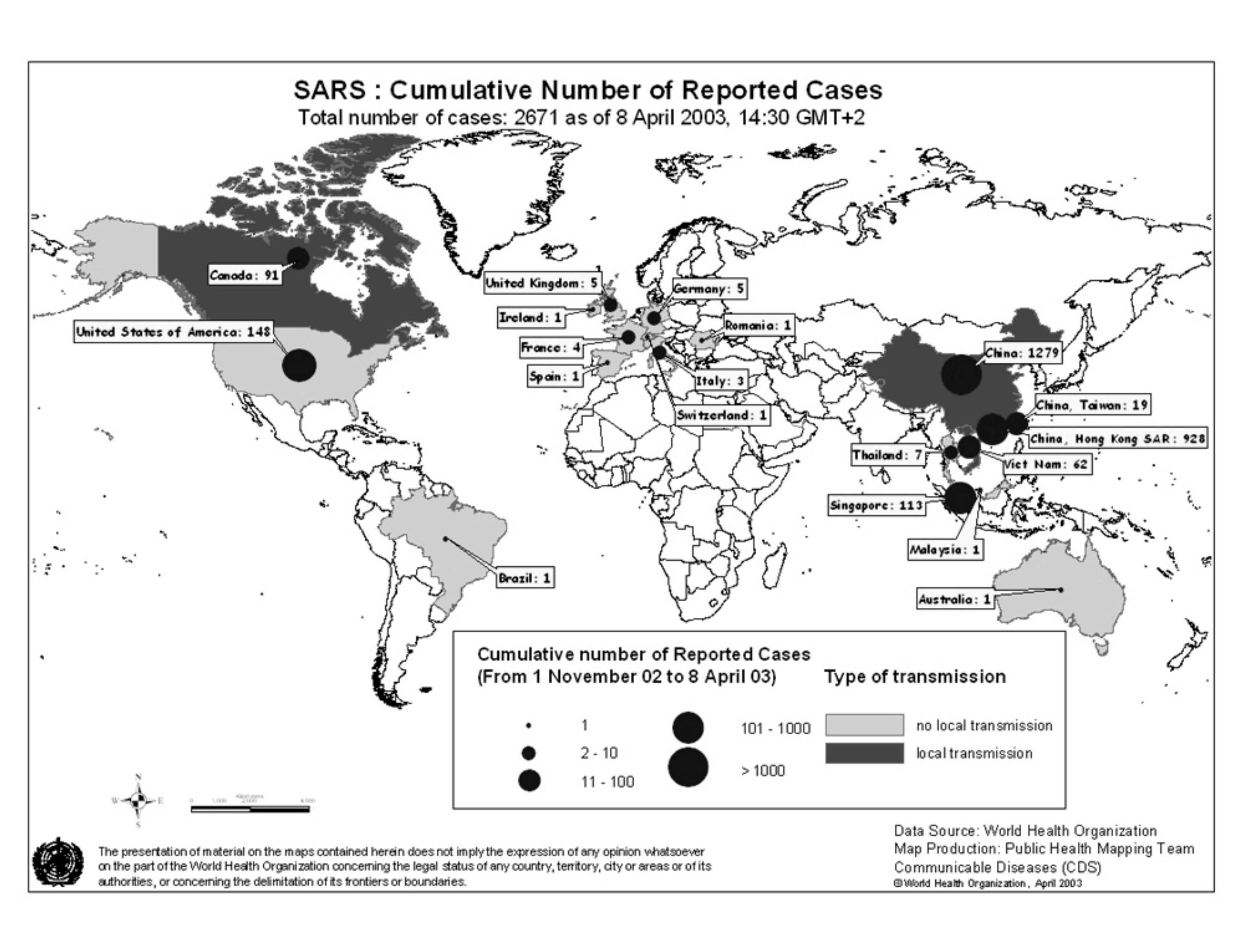}
\caption{SARS Map: Cumulative number of reported cases as of April 8, 2003 \cite{WorldHealthOrganizationSARS4}}
\label{SARS_Map}
\end{figure}

The SARS pandemic of 2002 is a useful case study for our modeling methodology.
The pandemic spread to 29 countries/regions in 2002 and 2003. In total 8,096 people were infected and 774 people died as of December 31, 2003 \cite{WorldHealthOrganizationSARS3}. Figure~\ref{SARS_Map} shows the spread of SARS as of April 8, 2003 \cite{WorldHealthOrganizationSARS4}. In 20 of 29 countries/regions, 100\% of total cases in the country were ``imported'' (as defined by WHO) from other countries \cite{WorldHealthOrganizationRegionalOffice2006}. 

The SARS pandemic is a particularly useful case study because we have high-fidelity data on the outbreak. First, the beginning and end of the pandemic are clear. According to the WHO, the first case was a male in his 40's in Guangdong, China, in November 2002. SARS started substantially spreading from Hong Kong to other countries in February 2003, infecting 29 countries and regions by July 2003. After that, there were no new cases except one infection through a laboratory accident. Second, the number of cases is clearly reported (and relatively small). WHO reported the cumulative number of cases and the number of new infected cases from March 17th to July 11th 2003 \cite{WorldHealthOrganizationSARS2}. Third, the number of infected countries is clearly shown. There are 29 countries/regions which are infected by SARS by the end of 2003 \cite{WorldHealthOrganizationSARS3}. Thus we have good data on the progress of infection in different countries and regions.

%% file: model.tex
We first introduce the main methodologies for modeling the spread of infectious diseases before describing our approach in detail.

\subsection{Infectious Disease Models}
\subsubsection{SIR Model}
The classic SIR model, proposed by Kermack and McKendrick in 1927 \cite{Capasso1993}, posits three classes of agents; Susceptible, Infectious, and Removed. Susceptible agents (hereafter denoted $S$) are vulnerable to a disease and have the potential to be infected. Infectious agents ($I$) are currently infected and have the risk of infecting $S$. Removed ($R$) agents are removed from the system -- they are either dead or acquired immunity.

Thus $R$ is not infected again. $R$ is also called Recovered when we assume it is not dead. When $R$ is not dead but has instead acquired immunity, the total population, ($S + I + R$), is constant. The model assumes that agents in the set $S$ are sometimes infected by a contact in $I$ and change to $R$ at a constant rate. This yields the expressions below for the transition of populations of these three classes.
\begin{equation}
\begin{cases}
\; \dfrac{dS}{dt}= -\beta S I \\ \\
\; \dfrac{dI}{dt}= \beta S I - \lambda I \\ \\
\; \dfrac{dR}{dt}=\lambda I 
\label{eq: SIR}
\end{cases}  
\end{equation}
where $\beta$ is the rate of infection from $S$ to $I$ and $\lambda$ is the rate of recovery from $I$ to $R$. $\lambda$ is inversely proportion to the average infectious period, $\tau$ : $\lambda = \tau^{-1}$. When we assume that the population is constant in this case, the total population $N$ is given by $S + I + R$. When $\beta / \lambda > 1$, the infection spreads since the probability that $S$ becomes $I$ is greater than the speed that $I$ becomes $R$. The \emph{basic reproduction number}, $R_0$, is the average number of persons infected by a single infected person when the population has no immunity and no control against the infection \cite{Yamamoto2006}. In the SIR differential equation model, the basic reproduction number is given by $R_0 = N\beta / \lambda$. If one infected person infects more than one susceptible person (i.e., $R_0 > 1$), secondary infection occurs and the infection spreads. On the other hand, if $R_0 \leq 1$, the disease converges in the system. Therefore $R_0 = 1$ is a threshold for spread.

\subsubsection{SEIR Model}
\label{SEIR Model}
The SEIR model is a derivative of the SIR model. SIR doesn't consider the incubation period. Thus, when $S$ is infected, it becomes $I$ immediately and starts to infect other $S$ \cite{WolframMathworld}. In the real world, there is some duration between the time that a person is infected and the time that he/she starts infecting others. The SEIR model denotes agents in the incubation period as belonging to class $E$ (exposed) \cite{Hethcote1980}. The corresponding transition equations are:
\begin{equation}
\begin{cases}
\; \dfrac{dS}{dt}= -\beta S I \\ \\
\; \dfrac{dE}{dt}= \beta S I - \lambda E \\ \\
\; \dfrac{dI}{dt}=  \lambda E - \lambda I \\ \\
\; \dfrac{dR}{dt}= \lambda I 
\label{eq: SEIR}
\end{cases}  
\end{equation}

\subsubsection{Network- and Agent-Based Models}

Agent-based modeling provides an explicit, local method of understanding the spread of infection. It allows for fine-grained control over many aspects of the dynamic model of disease spread, including geographic factors and agent movements. For example, 
Carley, \emph{et al} \cite{Carley2003} simulate the spread of anthrax and Epstein \cite{Epstein2004} investigates the spread of smallpox with agent based models. Deguchi, \emph{et al} have developed an Agent Based simulation language called SOARS, Spot Oriented Agent Role Simulator \cite{Deguchi20061,Deguchi20062} for simulating the spread of disease considering modules such as human activities, opportunity for contact between people in a society, disease state, and intervention to control the spread.

Network-based models typically represent agents as nodes on graphs and allow the connectivity structure of the graph to determine the possible spread of disease.
For example, extending an SIR model to networks would involve allowing 
a susceptible vertex $S$ to be infected by an infectious vertex $I$ only if $S$ is adjacent to $I$. Network-based models are useful in that they can reflect social and economic networks. 
People's behaviors and social contacts build the network and the infection route is on the network. 

\subsection{Our Approach}

\begin{figure}
\centering
\includegraphics[angle=90,width=4in]{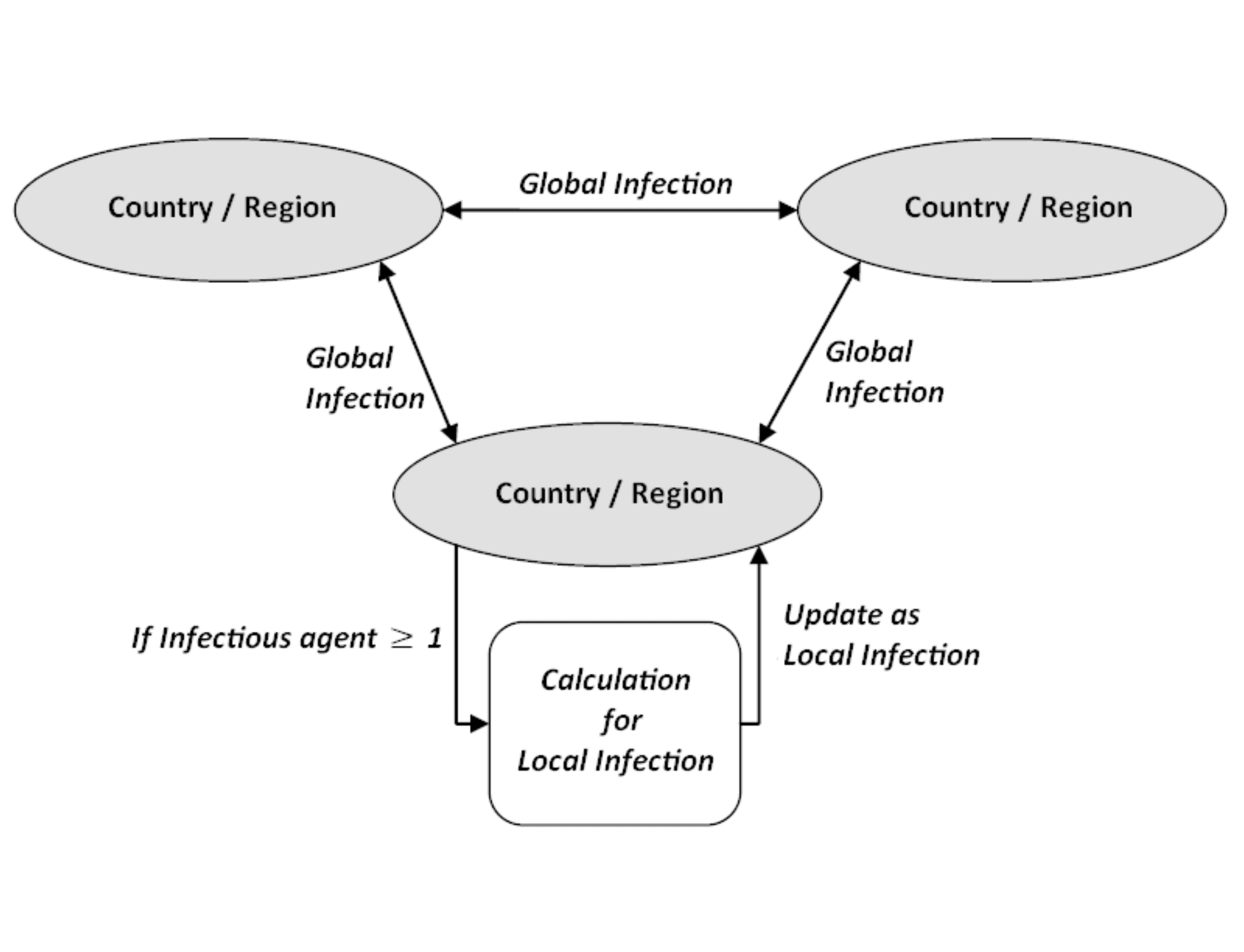}
\vspace{-.5in}
\caption{Structure of the simulation model}
\label{fig:Concept_Model}
\end{figure}

Our model uses local regions and interconnections between them. 
There are three possibilities for a new infection in a region; (1) infection from travelers from outside the region, (2) infection from returning travelers, and (3) infection from local persons. We denote infection types (1) and (2) as ``global'' infections and type (3) as ``local'' infections. Figure \ref{fig:Concept_Model} shows the basic structure of our model.

\paragraph{Global and Local Infections}
We assume that infection starts in a particular country or region and spreads from there. At each cycle, infections of all types can occur. Global infections (types 1 and 2) occur with frequencies that are dependent on the level of travel between regions, and local infections are mostly dependent on the population density of a region (details of the data used are below). 
Our local model is based on the concept behind the SEIR model. We consider the same four types of agents in each region: Susceptible, Exposed, Infectious, and Removed. When an infection occurs, agents are considered exposed. The model proceeds in time cycles $t$. The number of agents newly exposed in region $i$ at time $t$ through the global infection mechanism is modeled as $\text{EG}_{i}(t) = \sum_{j} I_{j}(t) \cdot T_{ij} \cdot P_{G}^*(t)$, where the sum is over other regions, $T_{ij}$ is the sum of travelers from region $i$ to region $j$ and the number of travelers from region $j$ to region $i$ (since infection can occur through both arriving and returning travelers), and $P_{G}^{*}(t)$ is a ``global infection coefficient'' at time $t$, described below.

Local infection follows a similar process, so that the number of agents newly exposed through the local infection mechanism at any time $t$ is given by $\text{EL}_{i}(t) = S_{i}(t) \cdot I_{i}(t) \cdot P_{L i}^*(t)$ where $P_{L i}^*(t)$ is a ``local infection coefficient'' (similar to the global infection coefficient, both are described in detail below).

It is assumed that agents go from exposed to infectious according to some incubation period that is disease-specific, and, similarly, from infectious to removed according to some disease-specific recovery period. For the purposes of this paper, we set these to 10 for both incubation period and infectious period, but these parameters can of course be varied for modeling other diseases.

\paragraph{Infection probabilities}

As awareness of a disease spreads, it is likely that heightened awareness and prevention measures start to reduce the spread of infection. We model this in our global and local infection coefficients, by introducing a term that dampens the coefficient over time. For global infection, we use $P_{G}^*(t) = P_{G}-(D_{G} \cdot t)$ where $P_{G}$ is a basic global infection coefficient, held constant across regions, and $D_{G}$ models the dampening effect.

We use a similar equation for the local infection coefficient, $P_{L i}^*(t) = P_{L i}-(D_{L} \cdot t)$. In this case, $D_{L}$ is assumed constant across regions, but the basic local infection coefficient $P_{L i}$ is region-specific, given by $P_{L i}=\rho_{i} \cdot C_{1} + C_{2}$ where $\rho_{i}$ is the population density of region $i$, assumed to be the primary driver of high local infection rates.

It is worth noting that the original SEIR model gives a similar type of equation for newly exposed agents $E = \beta \cdot S \cdot I$, where $\beta$ is the infection rate. The main novelty here is the combination of modeling a declining infection rate, and treating each region separately.

%% file: data.tex
There are several model parameters that need to be calibrated using real data. It is useful to consider some background information on the characteristics of SARS in this context.

\paragraph{Characteristics of SARS}
The SARS Coronavirus causes general infection with Viremia, especially severe pneumonia and intestine infection. It is transmitted primarily through droplet infection. Due to its resistance to dryness, it can also be transmitted through air. It is thought that the incubation period of SARS is usually 2-10 days and the average is 5 days \cite{Okada2003}. In the pandemic of 2002-2003, most countries reported a median incubation period of 4-5 days, and a mean of 4-6 days. In the incubation period, it is unlikely an infected person will spread the disease through droplet infection. The infectious period is thought to be about two weeks, with its peak from the 7th-10th day after infection \cite{Okada2003}. Transmission efficiency appears to be greatest from severely ill patients who are experiencing rapid clinical deterioration, usually during the second week of illness. Maximum virus excretion from the respiratory tract occurs on about day 10 of illness and then declines to 0\% by day 23. There are no reports on transmission beyond 10 days of fever resolution \cite{UniversityofHongKong}. The death rate varies by age group (SARS affects older patients much more severely), but the overall death rate was about 9.6\% in the 2002-2003 SARS pandemic, significantly higher than that of seasonal Influenzas. Another notable feature of SARS is that it is believed that ``super-spreading'' events, where a person infects many more than the average rate of infection, are a key component in its transmission. Our model does not deal explicitly with such levels of granularity, which may lead to some outlier predictions in areas where the law of large numbers does not take over. This is discussed further in Section \ref{section:discussion}

\subsection{Correlation between Pandemic and Traffic}
It is thought that the origin of SARS was Guangdong in China, quickly spreading to Hong Kong. Thus we consider countries/regions which have strong relationships with China and Hong Kong. At first we examine the numbers of travelers from China and Hong Kong and consider the ten countries/regions where the number of travelers to and from China and Hong Kong is the largest (see Table~\ref{Top10_from_to_China_HongKong}), yielding a total of 17 countries. 
16 of these 17 countries/regions were infected by SARS. Since there were 29 countries/regions in total with reported cases of SARS, half of them are represented in this table. Besides these 16 countries/regions, there are 13 other infected countries by SARS; Canada, France, India, Indonesia, Italy, Kuwait, New Zealand, Ireland, Romania, South Africa, Spain, Sweden, Switzerland.  We focus on these 30 countries/regions in our experiments. There are 8 countries/regions which had local spread: China, Hong Kong, Taiwan, Canada, Singapore, Vietnam, Philippines, and Mongolia. 7 of these 8  are included in Table~\ref{Top10_from_to_China_HongKong}. 

\begin{table}
\caption{Top 10 countries/regions in terms of number of travelers from/to China and Hong Kong in 2003 (dark-gray: country with local infection, light-gray: country with only imported cases, white: country without local Infection or imported cases, Created based on \cite{WorldTourismOrganization20041,WorldTourismOrganization20042})}
\label{Top10_from_to_China_HongKong}
\renewcommand{\arraystretch}{1.5}
{\fontsize{9pt}{8} \selectfont
\begin{center}						
\begin{tabularx}{150mm}{XrXr|XrXr}\toprule
\multicolumn{4}{c|}{From} &
\multicolumn{4}{c}{To} \\
\multicolumn{2}{c}{China} &
\multicolumn{2}{c|}{Hong Kong} &
\multicolumn{2}{c}{China} &
\multicolumn{2}{c}{Hong Kong} \\ \hline
\multicolumn{1}{l}{To}& 
\multicolumn{1}{l}{No.}& 
\multicolumn{1}{l}{To}& 
\multicolumn{1}{l|}{No.}& 
\multicolumn{1}{l}{From}& 
\multicolumn{1}{l}{No.}& 
\multicolumn{1}{l}{From}& 
\multicolumn{1}{l}{No.} \\ \hline

\multicolumn{1}{>{\columncolor[gray]{0.6}}X}{Hong Kong}	&	5,692,500	&	\multicolumn{1}{>{\columncolor[gray]{0.6}}X}{China}	&	58,770,063	&	\multicolumn{1}{>{\columncolor[gray]{0.6}}X}{Hong Kong}	&	58,770,063	&	\multicolumn{1}{>{\columncolor[gray]{0.6}}X}{China}	&	5,692,500	\\
\multicolumn{1}{>{\columncolor[gray]{0.8}}X}{Macao}	&	1,431,294	&	\multicolumn{1}{>{\columncolor[gray]{0.8}}X}{Macao}	&	1,218,648	&	\multicolumn{1}{>{\columncolor[gray]{0.8}}X}{Macao}	&	18,757,267	&	Japan	&	563,300	\\
\multicolumn{1}{>{\columncolor[gray]{0.6}}X}{Vietnam}	&	693,423	&	\multicolumn{1}{>{\columncolor[gray]{0.8}}X}{Thailand}	&	649,920	&	\multicolumn{1}{>{\columncolor[gray]{0.6}}X}{Taiwan}	&	2,731,897	&	\multicolumn{1}{>{\columncolor[gray]{0.8}}X}{United States}	&	532,500	\\
\multicolumn{1}{>{\columncolor[gray]{0.8}}X}{Russian Federation}	&	679,608	&	\multicolumn{1}{>{\columncolor[gray]{0.6}}X}{Taiwan}	&	287,312	&	Japan	&	2,254,800	&	\multicolumn{1}{>{\columncolor[gray]{0.6}}X}{Taiwan}	&	407,100	\\
\multicolumn{1}{>{\columncolor[gray]{0.8}}X}{Thailand}	&	624,214	&	Japan	&	260,214	&	\multicolumn{1}{>{\columncolor[gray]{0.8}}X}{Korea, Republic of}	&	1,945,484	&	\multicolumn{1}{>{\columncolor[gray]{0.8}}X}{United Kingdom}	&	235,100	\\
\multicolumn{1}{>{\columncolor[gray]{0.6}}X}{Singapore}	&	568,510	&	\multicolumn{1}{>{\columncolor[gray]{0.6}}X}{Singapore}	&	226,260	&	\multicolumn{1}{>{\columncolor[gray]{0.8}}X}{Russian Federation}	&	1,380,650	&	\multicolumn{1}{>{\columncolor[gray]{0.8}}X}{Korea, Republic of}	&	225,200	\\
\multicolumn{1}{>{\columncolor[gray]{0.8}}X}{Korea, Republic of}	&	513,236	&	\multicolumn{1}{>{\columncolor[gray]{0.8}}X}{Korea, Republic of}	&	156,373	&	\multicolumn{1}{>{\columncolor[gray]{0.8}}X}{United States}	&	822,511	&	\multicolumn{1}{>{\columncolor[gray]{0.8}}X}{Australia}	&	196,900	\\
Japan	&	448,782	&	\multicolumn{1}{>{\columncolor[gray]{0.6}}X}{Philippines}	&	139,753	&	\multicolumn{1}{>{\columncolor[gray]{0.6}}X}{Philippines}	&	457,725	&	\multicolumn{1}{>{\columncolor[gray]{0.6}}X}{Singapore}	&	184,200	\\
\multicolumn{1}{>{\columncolor[gray]{0.8}}X}{Malaysia}	&	350,597	&	\multicolumn{1}{>{\columncolor[gray]{0.8}}X}{United Kingdom}	&	131,000	&	\multicolumn{1}{>{\columncolor[gray]{0.8}}X}{Malaysia}	&	430,137	&	\multicolumn{1}{>{\columncolor[gray]{0.6}}X}{Philippines}	&	178,700	\\
\multicolumn{1}{>{\columncolor[gray]{0.8}}X}{Germany}	&	268,057	&	\multicolumn{1}{>{\columncolor[gray]{0.8}}X}{Australia}	&	129,292	&	\multicolumn{1}{>{\columncolor[gray]{0.6}}X}{Mongolia}	&	418,257	&	\multicolumn{1}{>{\columncolor[gray]{0.8}}X}{Macao}	&	156,100	\\
\bottomrule
\end{tabularx}						
\end{center}
}
\end{table}

\subsection{Correlation between Local Infection and Population Density}
\begin{figure}
\centering
\includegraphics[width=4in,angle=90]{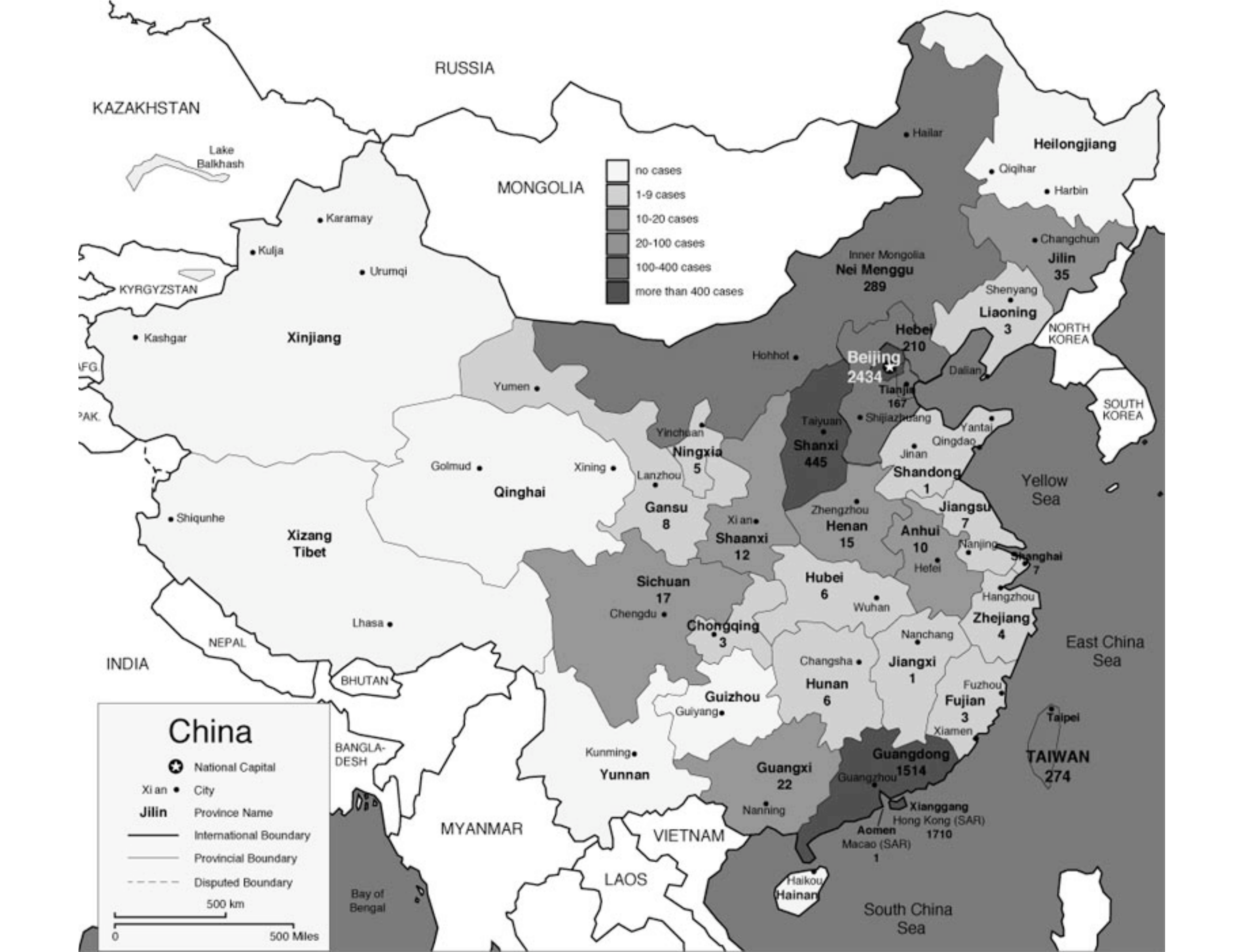}
\caption{Map of SARS cases by province in China as of May 18, 2003 \cite{NationsOnlineProject}}
\label{SARS_Map_China}
\end{figure}

We hypothesize that population density of an area is positively correlated with the local infection rate, because higher population densities lead to more frequent contact. We test this hypothesis using data from Chinese provinces, Hong Kong, and Taiwan, the most significant infected regions. Figure~\ref{SARS_Map_China} shows how the number of infections in different Chinese provinces varied greatly at the peak of the infection (from~\cite{NationsOnlineProject}), which makes it necessary to treat the individual regions separately. Since 97\% of infections occur in 6 provinces, we use data from these 6. They are 
Guangdong Province (the initial infected province), Beijing Municipality, Shanxi Province, Inner Mongolia Autonomous Region, Hebei Province, and Tianjin Municipality. Table \ref{PAD_8} shows basic data on population and density for each of the provinces, Hong Kong, and Taiwan. Using these 6 provinces, Hong Kong, and Taiwan, we can reject the null hypothesis that there is no correlation between population density and infection rate at the $0.01$-level. We choose the values for C1 and C2 by trial and error.

\begin{table}
\caption{Population, area, and population density in 6 provinces in mainland China, Hong Kong, and Taiwan \cite{NationalBureauofStatisticsofChina2005,UnitedNationsPopulationDivision,UnitedNationsStatisticsDivision}}
\label{PAD_8}
\renewcommand{\arraystretch}{1.5}
\scriptsize{
\begin{center}						
\begin{tabularx}{110mm}{Xrrr}\toprule
\multicolumn{1}{c}{} &
\multicolumn{1}{c}{Population} &
\multicolumn{1}{c}{Area} &
\multicolumn{1}{c}{Population Density} \\
\multicolumn{1}{c}{} &
\multicolumn{1}{c}{} &
\multicolumn{1}{c}{(sq. km)} &
\multicolumn{1}{c}{(per sq. km)} \\ \midrule
Beijing	&	17,422,637	&	16,801	&	1,037	\\
Guangdong	&	83,079,300	&	177,900	&	467	\\
Hebei	&	68,135,100	&	187,700	&	363	\\
Hong Kong	&	6,708,940	&	1,108	&	6,055	\\
Inner Mongolia	&	23,660,000	&	1,183,000	&	20	\\
Shanxi	&	33,398,400	&	156,800	&	213	\\
Taiwan	&	23,067,604	&	36,006	&	641	\\
Tianjin	&	11,760,000	&	11,760	&	1,000	\\
\bottomrule
\end{tabularx}						
\end{center}
}
\end{table}

\subsection{Passenger Traffic}

Our initial simulations are focused on the 6 regions of China, Taiwan, and Hong Kong. 
Table~\ref{Traffic_3} shows the number of travelers among the three countries. However, it is difficult to estimate travel between the regions of China, or to allocate travelers from China to the other countries amongst the regions of China.

\begin{table}
\caption{Number of travelers between the three regions in 2003 \cite{WorldTourismOrganization20041,WorldTourismOrganization20042}}
\label{Traffic_3}
\renewcommand{\arraystretch}{1.5}
\scriptsize{
\begin{center}						
\begin{tabularx}{75mm}{l|rrr}\toprule
\multicolumn{1}{c|}{} &
\multicolumn{1}{c}{China} &
\multicolumn{1}{c}{Hong Kong} &
\multicolumn{1}{c}{Taiwan} \\ \hline
China	&	-	&	5,692,500	&	NA*	\\
Hong Kong	&	58,770,063	&	-	&	287,312	\\
Taiwan	&	2,731,897	&	407,100	&	-	\\
\bottomrule
\multicolumn{4}{l}{Row: Origin, Column: Destination} \\
\multicolumn{4}{X}{*Civil travel from China to Taiwan was not permitted in 2003 (Lifted on July 18, 2008)} \\
\end{tabularx}						
\end{center}
}
\end{table}

In order to approximate this travel information, we use data on passenger land traffic and civil aviation in China in 2007~\cite{NationalBureauofStatisticsofChina2007}. Table~\ref{Passenger_Traffic_Total} shows the each passenger traffic and the total \cite{NationalBureauofStatisticsofChina2007}. We use land traffic data for the 6 provinces we are interested in, as shown in Table~\ref{Passenger_Traffic_Region} \cite{NationalBureauofStatisticsofChina2007}.  We compute the share of each region in the total, where the total share is 100. 

Then, based on Tables~\ref{Passenger_Traffic_Total} and \ref{Passenger_Traffic_Region}, we approximate the number of travelers between two regions by assuming that the share of a region is directly proportional to the number of travelers to the region. Also we assume that the share of passenger traffic by air is proportional to the share of passenger traffic by land. 

\begin{table}
\caption{Total passenger traffic in China in 2007 \cite{NationalBureauofStatisticsofChina2007}}
\label{Passenger_Traffic_Total}
\renewcommand{\arraystretch}{1.5}
\scriptsize{
\begin{center}						
\begin{tabularx}{140mm}{Xrrrrr}\toprule
\multicolumn{1}{c}{} &
\multicolumn{1}{c}{Railways} &
\multicolumn{1}{c}{Highways} &
\multicolumn{1}{c}{Waterways} &
\multicolumn{1}{c}{Civil Aviation} &
\multicolumn{1}{X}{Total (Total of Land)} \\ \midrule
  Passengers &	135,670 	&	2,050,680 	&	22,835 	&	18,576 	&	2,227,761 (2,209,185)	\\
(10,000 persons) & & & & & \\
  Share in Total (\%)	&	6.09 	&	92.05 	&	1.03 	&	0.83 	&	100 (99.17) 	\\
\bottomrule
\end{tabularx}						
\end{center}
}
\end{table}

\begin{table}
\caption{Passenger traffic by region in China in 2007 (civil aviation traffic is not included) \cite{NationalBureauofStatisticsofChina2007}}
\label{Passenger_Traffic_Region}
\renewcommand{\arraystretch}{1.5}
\scriptsize{
\begin{center}						
\begin{tabularx}{80mm}{Xrr}\toprule
\multicolumn{1}{c}{} &
\multicolumn{1}{c}{Passengers} &
\multicolumn{1}{c}{Share in Total (\%)} \\
\multicolumn{1}{c}{} &
\multicolumn{1}{c}{(10,000 persons)} &
\multicolumn{1}{c}{} \\
 \midrule
  Beijing	&	16,190 	&	0.732849005	\\
  Tianjin	&	6,829 	&	0.309118965	\\
  Hebei	&	88,886 	&	4.023496235	\\
  Shanxi	&	43,866 	&	1.985620174	\\
  Inner Mongolia	&	38,678 	&	1.750778371	\\
  Guangdong	&	199,162 	&	9.015179308	\\
  Others	&	1,815,574 	&	82.18295794	\\ \hline
  National Total	&	2,209,185 	&	100	\\
\bottomrule
\end{tabularx}						
\end{center}
}
\end{table}

We estimate travel between the different regions of China and Hong Kong and Taiwan by using the share of the airport of each region in China in the national total. Table~\ref{Traffic_Air_China} shows the number of passengers using the main airport in 6 regions of China and the share in the national total \cite{CivilAviationAdministrationofChina}. We apportion the number of travelers between China and Hong Kong or Taiwan according to the share. For example, the share of Beijing airport in the national total is 13.7859\%. The number of travelers from China to Hong Kong is 5,692,500. The number of travelers from Hong Kong to China is 58,770,063. Thus the total number of travelers between China and Hong Kong is 64,462,563. The number of travelers between Beijing to Hong Kong is obtained as; 0.137859 $\times$ 64,462,563 $\approx$ 8,886,700. 

\begin{table}
\caption{Passengers passing through the main airport in 6 regions of China in 2007 \cite{CivilAviationAdministrationofChina}}
\label{Traffic_Air_China}
\renewcommand{\arraystretch}{1.5}
\scriptsize{
\begin{center}						
\begin{tabularx}{80mm}{Xrr}\toprule
\multicolumn{1}{c}{} &
\multicolumn{1}{c}{Passengers} &
\multicolumn{1}{c}{Share in Total (\%)} \\
 \midrule
Beijing	&	55,938,136 	&	13.7859	\\
Tianjin	&	4,637,299 	&	1.1429	\\
Hebei	&	1,043,688 	&	0.2572	\\
Shanxi	&	4,312,910 	&	1.0629	\\
Inner Mongolia	&	2,121,905 	&	0.5229	\\
Guangdong *	&	54,835,981 	&	13.5143	\\
Others	&	282,872,185 	&	69.7138	\\ \hline
National Total	&	405,762,104 	&	100	\\
\bottomrule
\multicolumn{3}{l}{*Including both Guangzhou airport and Shenzhen airport} \\
\end{tabularx}						
\end{center}
}
\end{table}

%% file: results.tex
\subsection{Results for China, Hong Kong, and Taiwan}

For the preliminary experiment, we simulate with 6 regions in the Chinese mainland, Hong Kong, and Taiwan. The number of susceptible agents in each region/country is initially equal to the population of each country.  Table~\ref{Parameter_SARS} shows the summary for parameter values used in simulation. These simulation parameters were chosen to provide a good fit to data from this initial simulation, but we discuss below several inferences that can be made because many of the parameters are constant, exploiting the granularity of the model. Then, in the second part of this section, we use the same parameters to extend the model to 30 countries/regions, which provides a test for the parameters, allowing us to evaluate the benefits and drawbacks in a validation setting.

\begin{table}
\caption{Parameter values in simulation}
\label{Parameter_SARS}
\renewcommand{\arraystretch}{1.5}
{\fontsize{8pt}{7} \selectfont
\begin{center}						
\begin{tabularx}{80mm}{lX}\toprule
Parameter & Value \\ \midrule
$P_{G}$ & $2.0 \times 10^{-7} $ \\
$P_{L i}$ & Depends on $Density_{i}$, $C_{1}$, and $C_{2}$ \\
$D_{G}$ &  $5.0 \times 10^{-9} $\\ 
$D_{L}$ &  $2.5 \times 10^{-7} $\\
$C_{1}$ & $7.23 \times 10^{-9}$ \\
$C_{2}$ & $7.69 \times 10^{-6}$ \\
$Incubation\_Period$ & 10\\
$Infectious\_Period$ & 10\\
$Run\_Cycle$ & 100\\
$Density_{i}$ & See Table~\ref{PAD_8}\\
$Population_{i}$ & See Table~\ref{PAD_8}\\
$T_{ij}$ & See Table~\ref{Travelers_SARS} \\
\bottomrule
\end{tabularx}						
\end{center}
}
\end{table}

\begin{figure}[h]
\begin{minipage}[c]{0.48\textwidth}
\begin{center}
\includegraphics*[width=3in]{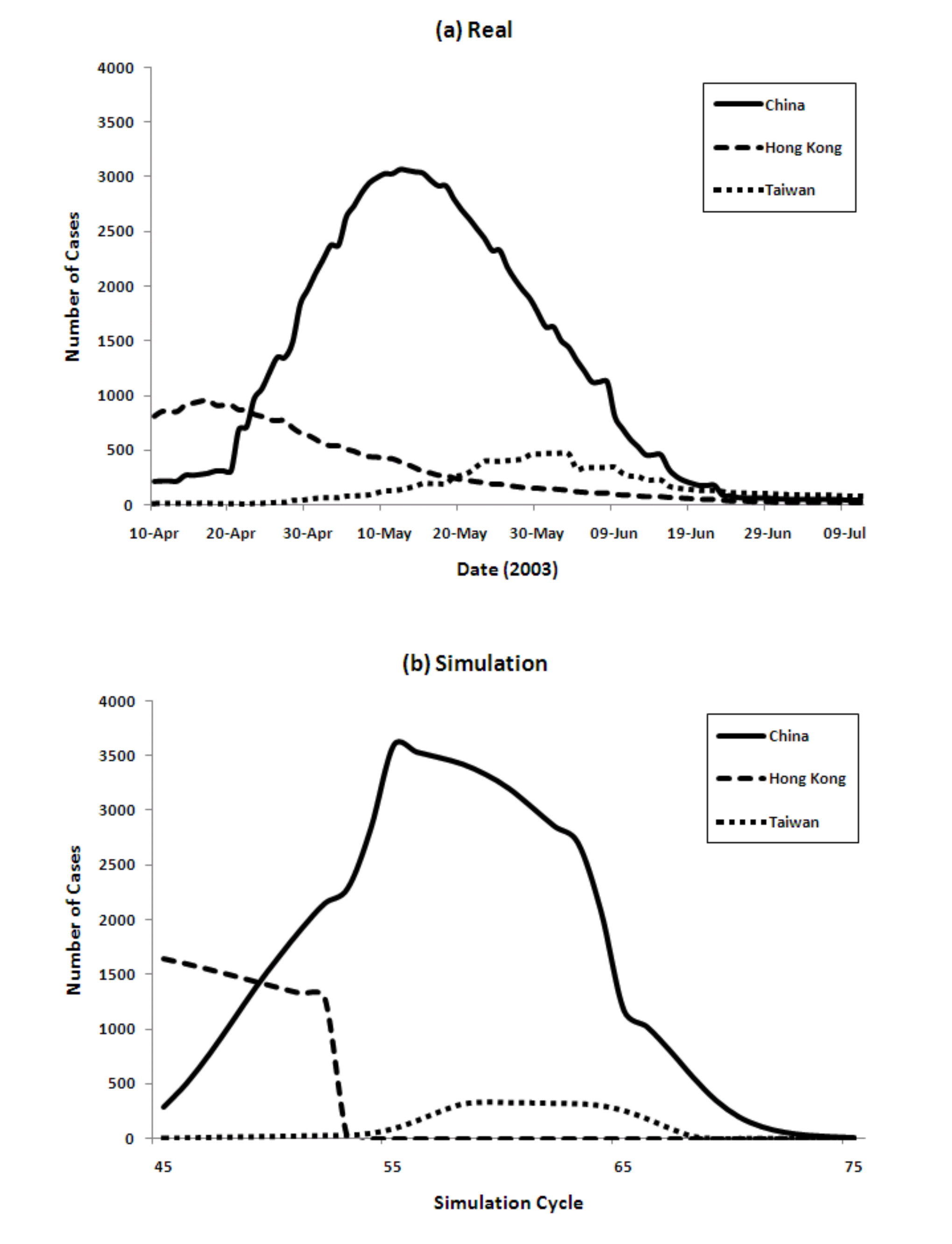}\\
\end{center}
\end{minipage}
\begin{minipage}[c]{0.48\textwidth}
\begin{center}
\includegraphics*[width=3in]{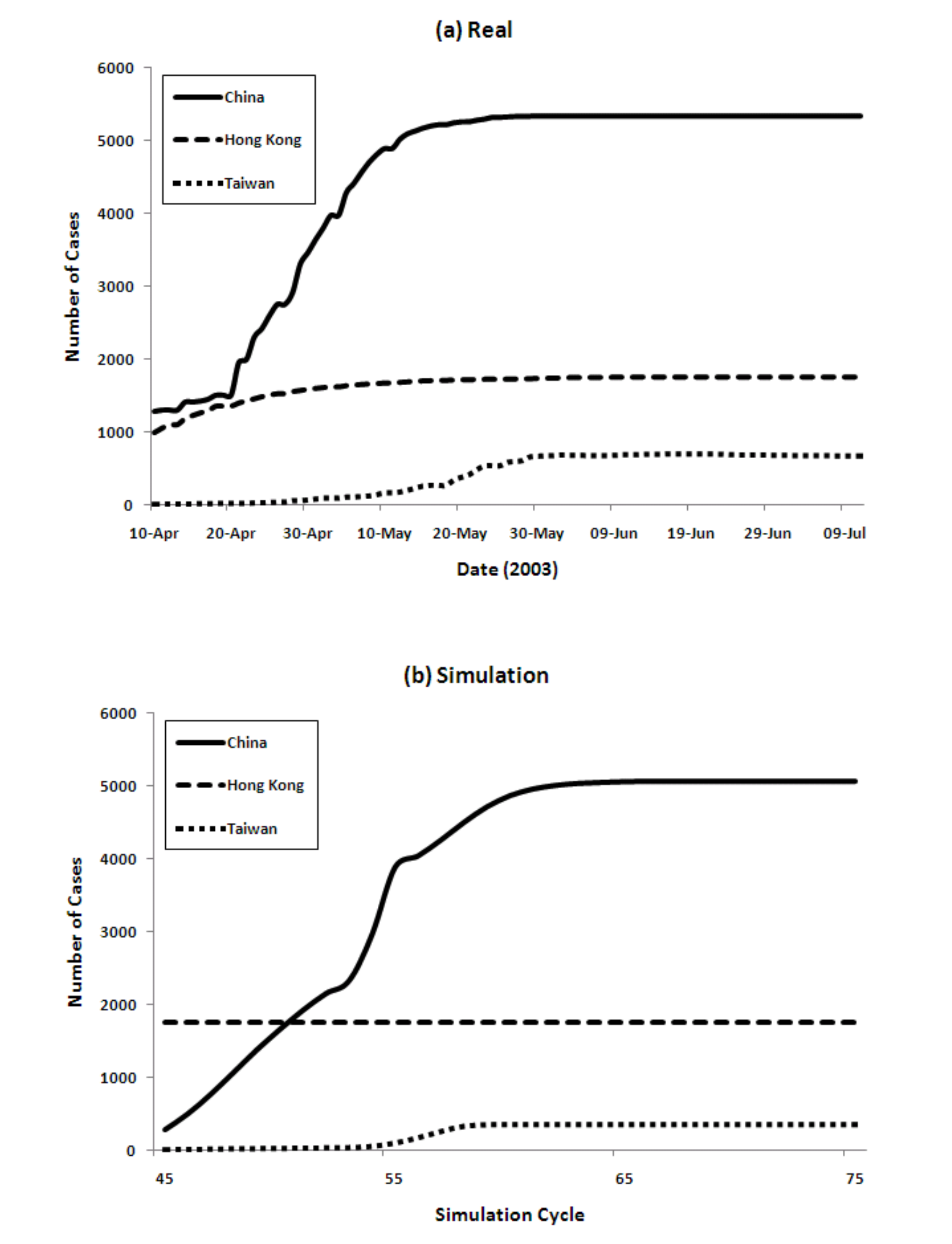}\\
\end{center}
\end{minipage}
\caption{Three country model for dynamics of the spread of SARS (left) and the cumulative number of cases (right), comparing reality and model predictions
\label{fig:InfectedCumSARS3}}
\end{figure}

Figure~\ref{fig:InfectedCumSARS3} show the transition of the number of infected cases and the number of cumulative cases respectively, comparing real data and the results of our model. For the model we show data from time cycles 45 through 75. The results of China's 6 regions are summed up and the total is shown for China. 

The figure shows that our model captures both the dynamics of the spread of SARS, as well as the total numbers, very well. The peaks come in order: Hong Kong, China, and Taiwan. The  model achieves this without using any special parameters that vary in different countries.
Populations, densities, and travel data are all taken from the real world. The SARS epidemic started spreading from Hong Kong and immediately reached mainland China. The peak of Hong Kong comes earlier than that of China since the population density is higher. However, the curve decreases from some point because the percentage of susceptible agents in the population decreases and the percentage of Recovered agents increases. Then the number of Infectious agents decreases. After that, the number of infected agents in China increases. Because of its population, the number of infected agents at its peak in China is the largest among the three countries/regions. The peak in Taiwan is slightly delayed because of the time lag in the infection reaching Taiwan.

\paragraph{Region-wise Breakdown}

\begin{figure}
\centering
\includegraphics[width=0.75\textwidth,angle=90]{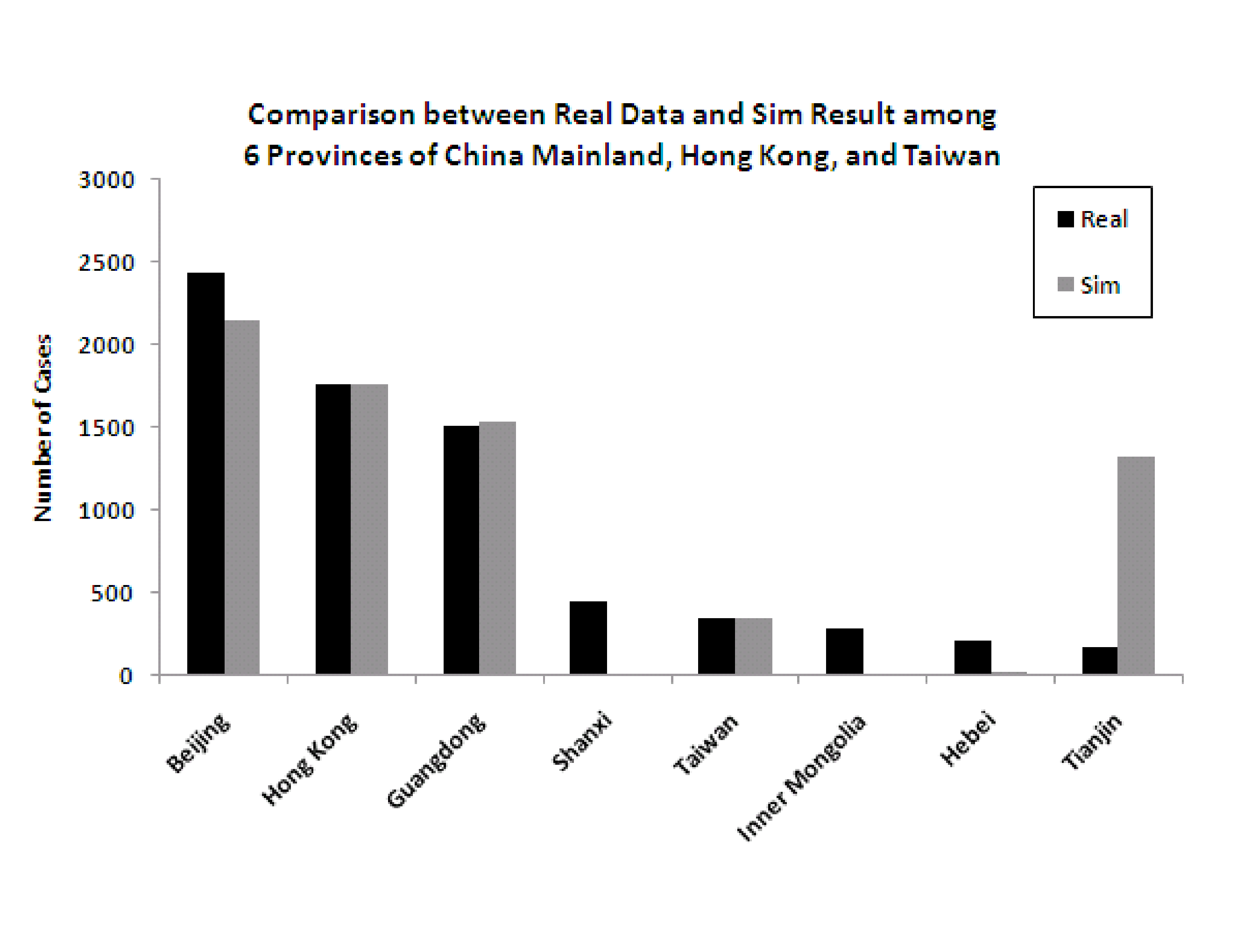}
\vspace{-.5in}
\caption{Total cases predicted in simulation and in reality for the eight modeled regions
\label{Comp_Case_SARS_Provinces}}
\end{figure}

\begin{figure}
\begin{minipage}[c]{0.48\textwidth}
\begin{center}
\includegraphics*[width=2.5in,angle=90]{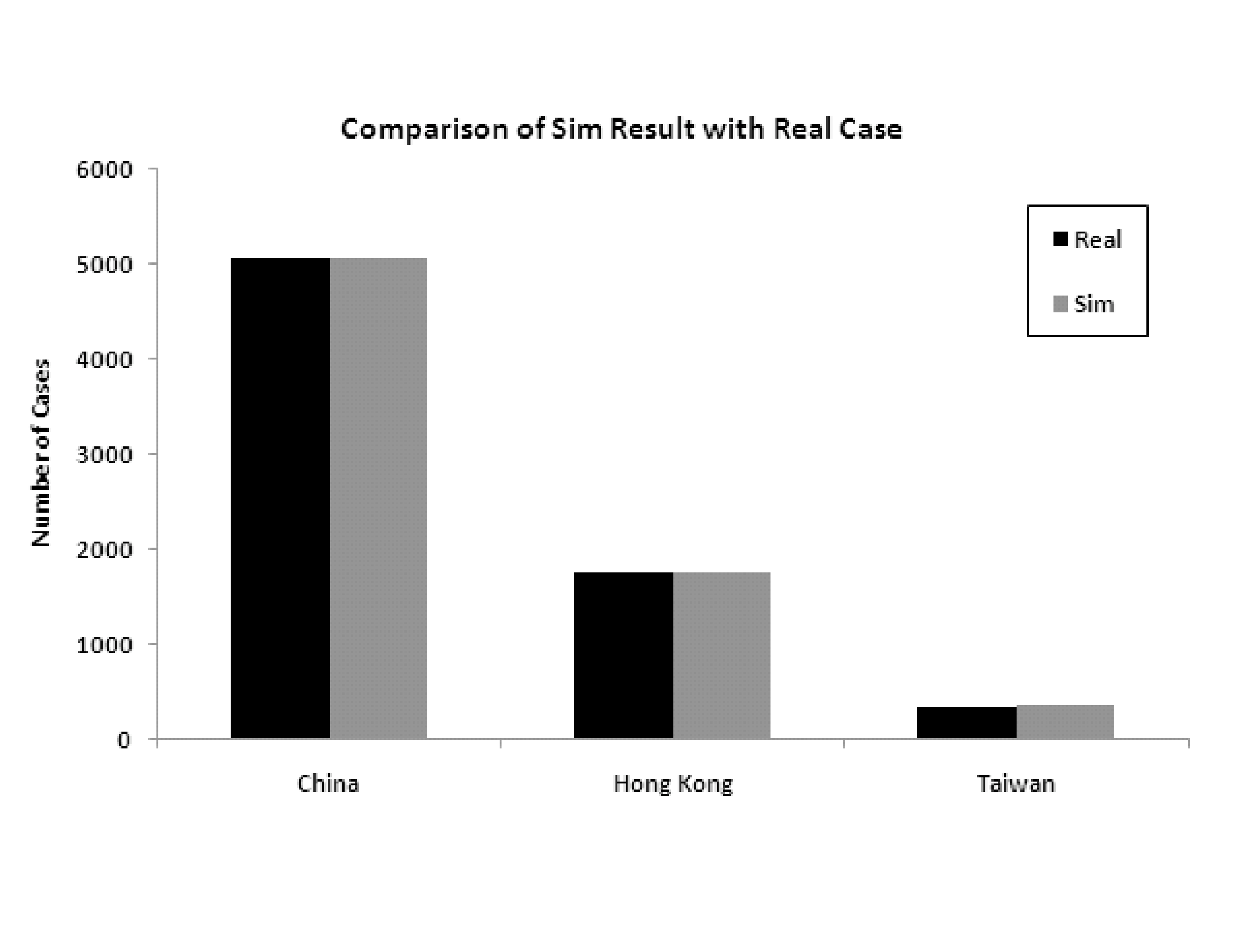}\\
(a) Modeling the 6 provinces independently
\end{center}
\end{minipage}
\begin{minipage}[c]{0.48\textwidth}
\begin{center}
\includegraphics*[width=2.5in,angle=90]{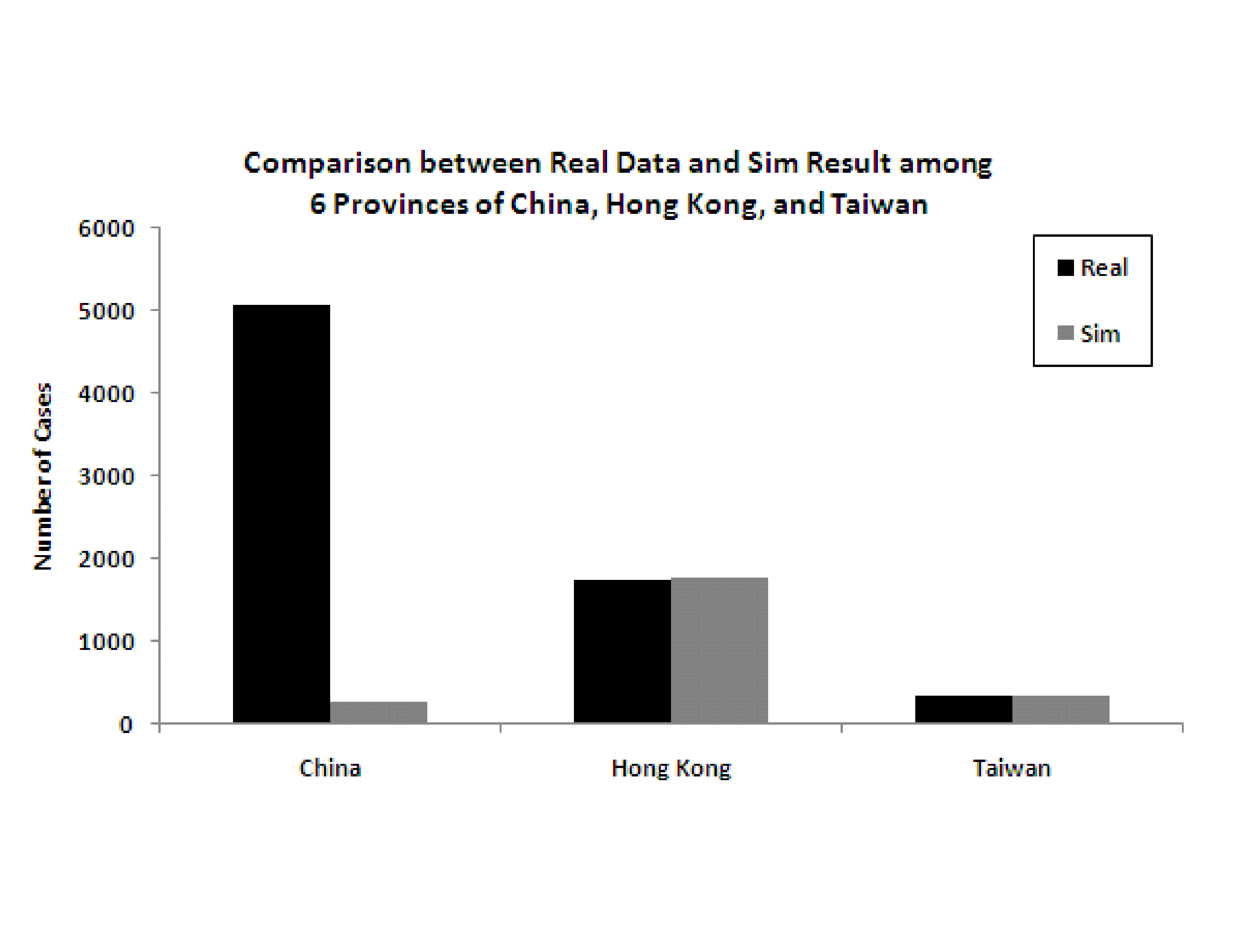}\\
(b) Modeling the 6 provinces as an aggregate
\end{center}
\end{minipage}
\caption{Model predictions for total infection in China, Hong Kong and Taiwan when splitting the 6 infected provinces versus aggregating them into one for modeling
\label{fig:allChinaVsSixProv}}
\end{figure}

Figure~\ref{Comp_Case_SARS_Provinces} shows the predicted (from the model) and actual number of cases for each of the eight modeled regions. While the fit is good for several of the most important (in terms of number of cases) regions, and therefore the overall numbers are good, there are some discrepancies for some of the regions that had a relatively fewer number of cases. Specifically, the model underpredicts the number of cases for some of the less densely populated provinces of China (Shanxi, Inner Mongolia, and Hebei) and overpredicts for one of the more densely populated regions (Tianjin). There are idiosyncratic events associated with the spread of any pandemic, so it is not entirely surprising that some of the results do not match perfectly. The next section considers anomalies in more detail, where some data is available. But it is important to note that the level of granularity in modeling is very important. Figure \ref{fig:allChinaVsSixProv} shows the difference in the model in two cases: one where the six infected provinces in China are modeled independently, and one where the six provinces are aggregated into one, using aggregated data on population density, travel etc.
The figure clearly shows that the more granular model is a much better fit to the data.

\subsection{Modeling 30 Countries/Regions}

As mentioned above, we use the parameters from the 8 region/country simulation to extend the model to 30 total countries (35 region/countries, since we continue to divide China into 6 regions). Again, we use real population, density, and international travel data from the 27 new countries (for Canada and Vietnam we use only Toronto and Hanoi, since only these regions had local spread cases \cite{WorldHealthOrganizationRegionalOffice2006}). Table~\ref{Travelers_SARS} show the expected number of travelers between countries/regions.
We again apportion the number of travelers between each region in mainland China and other countries based on the share of each region. 

\begin{table}
\caption{Expected number of travelers between countries/regions, 2004 (from \cite{WorldTourismOrganization20041}\cite{WorldTourismOrganization20042})}
\label{Travelers_SARS}
\renewcommand{\arraystretch}{1.5}
{\fontsize{6pt}{6} \selectfont
\begin{center}						
\begin{tabular}{l|rrrrrrr}\toprule
\multicolumn{1}{c|}{} &
\multicolumn{7}{c}{Origin} \\ 
\multicolumn{1}{c|}{Destination} &
\multicolumn{1}{c}{Beijing} &
\multicolumn{1}{c}{Tianjin} &
\multicolumn{1}{c}{Hebei} &
\multicolumn{1}{c}{Shanxi} &
\multicolumn{1}{c}{Inner Mongolia} &
\multicolumn{1}{c}{Guangdong} &
\multicolumn{1}{c}{Hong Kong} \\ \hline
Beijing	&	0	&	1,009,387	&	13,140,321	&	6,484,244	&	5,717,286	&	29,449,654	&	8,886,773	\\
Tianjin	&	1,009,387	&	0	&	5,542,547	&	2,735,034	&	2,411,533	&	12,421,773	&	736,718	\\
Hebei	&	13,140,321	&	5,542,547	&	0	&	35,604,996	&	31,393,629	&	161,708,110	&	165,808	\\
Shanxi	&	6,484,244	&	2,735,034	&	35,604,996	&	0	&	15,491,550	&	79,796,745	&	685,183	\\
Inner Mongolia	&	5,717,286	&	2,411,533	&	31,393,629	&	15,491,550	&	0	&	70,358,368	&	337,103	\\
Guangdong	&	29,449,654	&	12,421,773	&	161,708,110	&	79,796,745	&	70,358,368	&	0	&	8,711,676	\\
Hong Kong	&	8,886,773	&	736,718	&	165,808	&	685,183	&	337,103	&	8,711,676	&	0	\\
Taiwan	&	376,618	&	31,222	&	7,027	&	29,038	&	14,286	&	369,197	&	694,412	\\
Australia	&	58,114	&	4,818	&	1,084	&	4,481	&	2,204	&	56,969	&	326,192	\\
Canada	&	42,295	&	3,506	&	789	&	3,261	&	1,604	&	41,462	&	233,432	\\
France	&	127,391	&	10,561	&	2,377	&	9,822	&	4,832	&	124,881	&	64,800	\\
Germany	&	67,562	&	5,601	&	1,261	&	5,209	&	2,563	&	66,231	&	88,100	\\
India	&	33,121	&	2,746	&	618	&	2,554	&	1,256	&	32,468	&	114,770	\\
Indonesia	&	37,595	&	3,117	&	701	&	2,899	&	1,426	&	36,855	&	205,328	\\
Ireland, Republic of	&	1,746	&	145	&	33	&	135	&	66	&	1,712	&	0	\\
Italy	&	26,404	&	2,189	&	493	&	2,036	&	1,002	&	25,884	&	571,866	\\
Japan	&	372,714	&	30,898	&	6,954	&	28,737	&	14,138	&	365,371	&	823,514	\\
Korea, Republic of	&	338,958	&	28,100	&	6,324	&	26,134	&	12,858	&	332,279	&	381,573	\\
Kuwait	&	548	&	45	&	10	&	42	&	21	&	537	&	14	\\
Macao	&	2,783,184	&	230,727	&	51,928	&	214,587	&	105,575	&	2,728,346	&	1,374,748	\\
Malaysia	&	107,632	&	8,923	&	2,008	&	8,299	&	4,083	&	105,511	&	220,027	\\
Mongolia	&	70,114	&	5,813	&	1,308	&	5,406	&	2,660	&	68,733	&	380	\\
New Zealand	&	15,084	&	1,250	&	281	&	1,163	&	572	&	14,787	&	61,247	\\
Philippines	&	67,519	&	5,597	&	1,260	&	5,206	&	2,561	&	66,188	&	318,453	\\
Romania	&	2,345	&	194	&	44	&	181	&	89	&	2,299	&	0	\\
Russian Federation	&	284,026	&	23,546	&	5,299	&	21,899	&	10,774	&	278,430	&	3,585	\\
Singapore	&	130,496	&	10,818	&	2,435	&	10,061	&	4,950	&	127,924	&	410,460	\\
South Africa	&	8,472	&	702	&	158	&	653	&	321	&	8,305	&	18,600	\\
Spain	&	4,089	&	339	&	76	&	315	&	155	&	4,009	&	21,500	\\
Sweden	&	6,899	&	572	&	129	&	532	&	262	&	6,763	&	0	\\
Switzerland	&	11,890	&	986	&	222	&	917	&	451	&	11,655	&	45,642	\\
Thailand	&	124,024	&	10,282	&	2,314	&	9,562	&	4,705	&	121,581	&	790,020	\\
United Kingdom	&	49,123	&	4,072	&	917	&	3,787	&	1,863	&	48,155	&	366,100	\\
United States	&	135,080	&	11,198	&	2,520	&	10,415	&	5,124	&	132,418	&	646,612	\\
Vietnam	&	113,573	&	9,415	&	2,119	&	8,757	&	4,308	&	111,336	&	3,264	\\
\bottomrule
\end{tabular}						
\end{center}
}
\end{table}

\begin{figure}
\centering
\includegraphics[width=0.75\textwidth,angle=90]{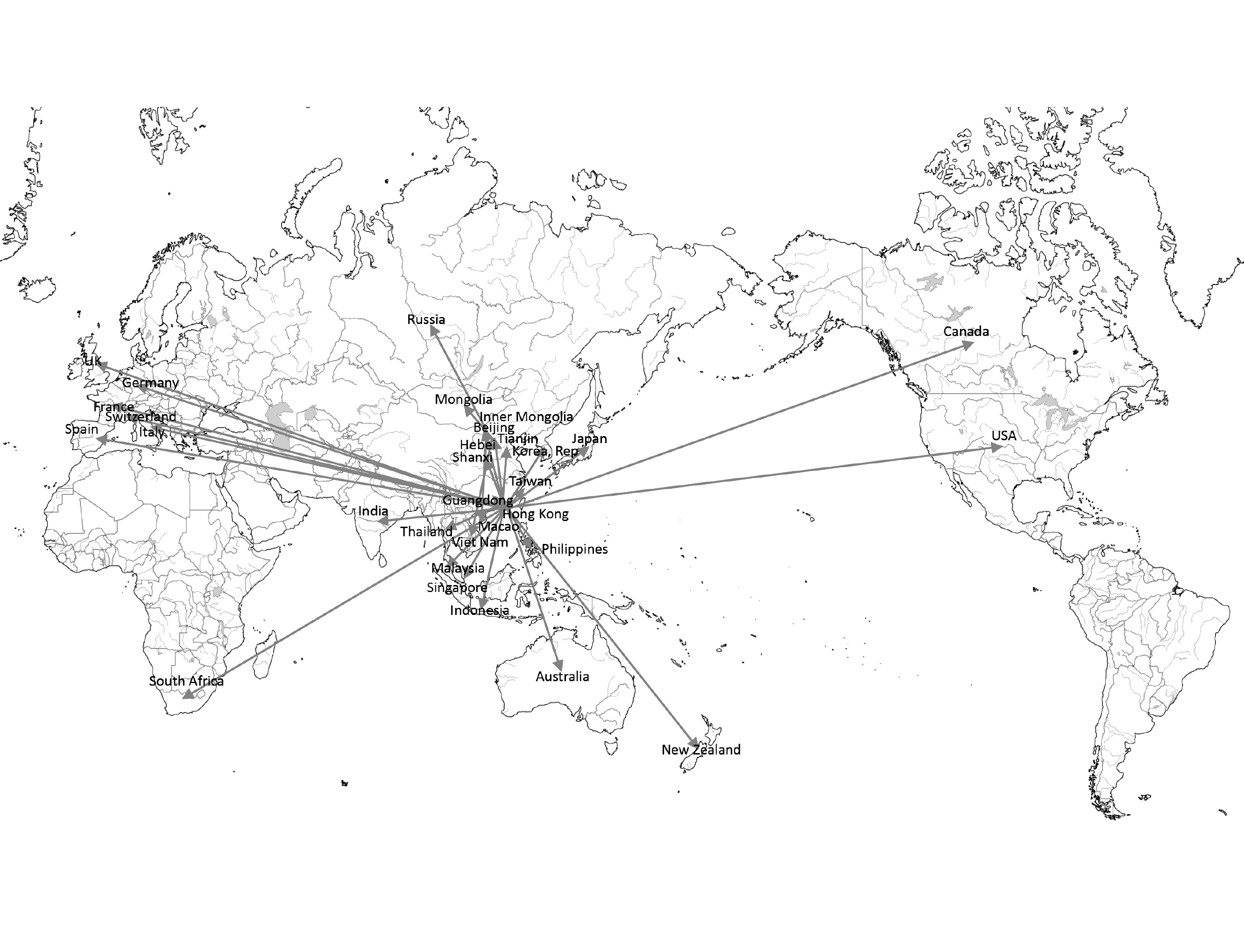}
\vspace{-.5in}
\caption{Infection route of SARS in simulation}
\label{Route_SARS}
\end{figure}

Figure~\ref{Route_SARS} shows the infection route in our model. Most countries are infected from Hong Kong or Guangdong. Some countries are infected from other regions. For example, Vietnam, Mongolia, and Russia are infected from Beijing. 

\begin{figure}
\centering
\includegraphics[width=0.75\textwidth]{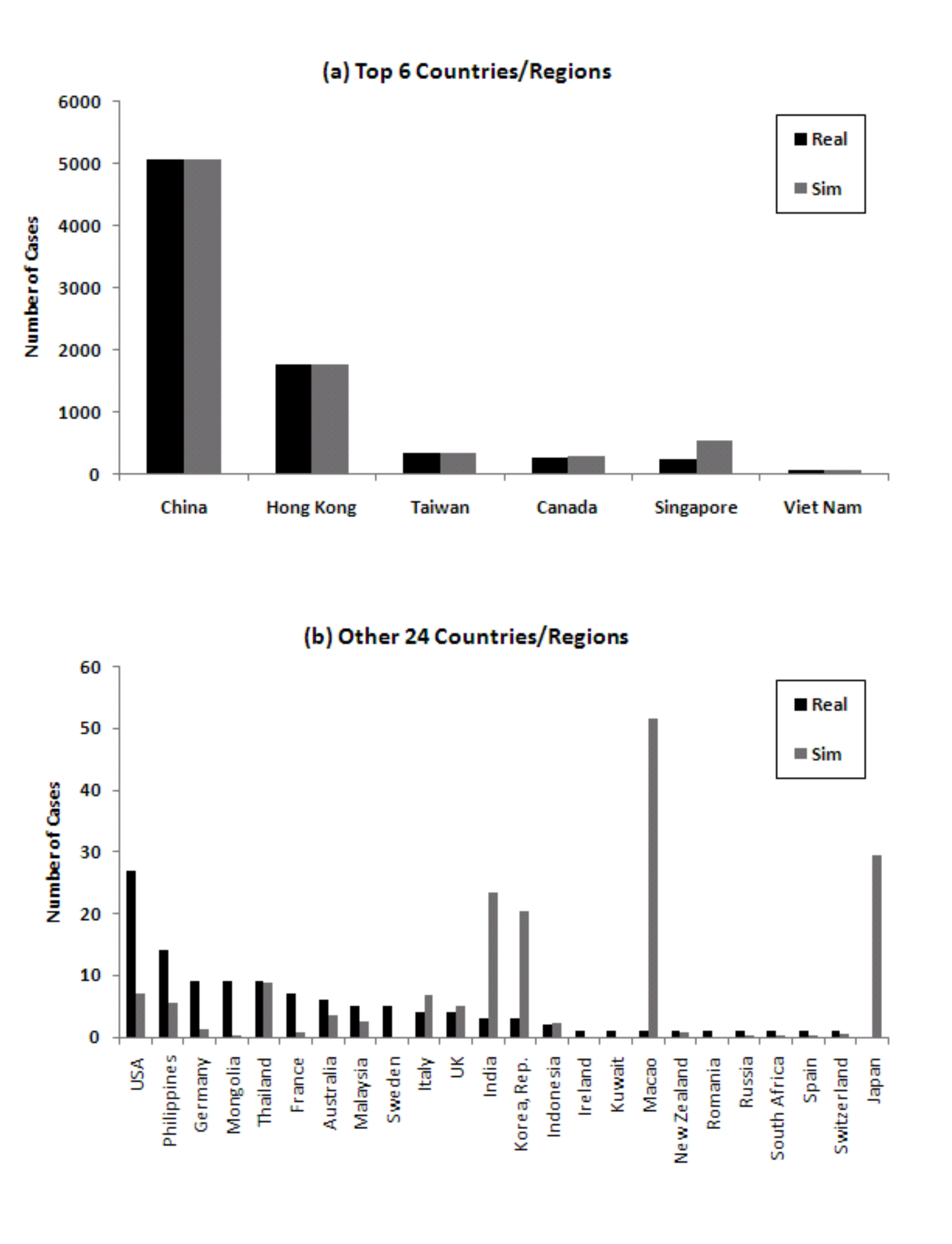}
\vspace{-.5in}
\caption{Comparison of number of cumulative cases in 30 countries/regions ((a) Top 6 countries/regions, (b) 24 other countries/regions). Note the different Y axes.}
\label{Comp_Case_SARS_30}
\end{figure}

Figure~\ref{Comp_Case_SARS_30} shows the comparison of the number of cumulative cases in simulation with real data. Especially for the significantly impacted countries, the number of cases corresponds well. In the real world, there were 8 countries/regions which had local spread. In the model, 18 countries/regions develop local spread. There are four true statistical outliers in the data in terms of number of cases predicted by the model versus number of cases experienced in reality. These are Singapore, Macao, Canada, and Japan. 

\paragraph{Discussion of anomalies}
We hypothesize that the outliers in this case are related to the nature of the spread of SARS. An early, chance outbreak, in a country or region can lead to significantly more cases than expected. Similarly, if a country manages to avoid a case of SARS for longer than predicted by international travel data, heightened awareness and prevention strategies will lead to many fewer cases than expected. For SARS in particular, this factor may be particularly important, because there is considerable evidence that some people infected with SARS are ``super spreaders'' who may affect the trajectory of the spread.  While an infected person infects, on average, 1-3 people \cite{Okada2003}, some infected people pass the virus to many other people \cite{WorldHealthOrganizationRegionalOffice2006}.  Although it is not clear what causes someone to become a super spreader, it is suspected that a person who has a chronic illness such as diabetes is more likely to be a super spreader \cite{Okada2003}. The origin of SARS is a case in point. A physician became ill on February 15th 2003 after caring for patients who had developed a strange new form of pneumonia in Guangdong. He stayed at the Metropole Hotel in Hong Kong on February 21st. On March 4th, he died of what would later be called SARS. During his one-night stay at the Metropole Hotel, the SARS virus had passed to at least 15 other guests at the hotel. The virus then spread around the world, leading to outbreaks in other countries \cite{WorldHealthOrganizationRegionalOffice2006}.

In each of the outlier cases, where the model makes a significantly different prediction than the actual trajectory of the pandemic, it turns our that the first reported case happened at a different time than would statistically be predicted by travel flows. While Macao, Japan, and Republic of Korea have large numbers of travelers from/to China and Hong Kong. these countries experienced much less infection than predicted by the model. It turns out that each of these countries experienced its first infection at a much later date than predicted, as shown in Table~\ref{Date_of_Onset}. Republic of Korea imported the first case in April 25th, Macao imported the first case in May 5th 2003, and Japan was never infected. These countries imported their first cases one or more months after Vietnam, Canada, Taiwan, Singapore and the Philippines. Meanwhile, Canada, despite being less strongly linked by travel to China and Hong Kong, was infected on February 23rd, early in the pandemic (in fact, from the original super-spread event at the Metropole Hotel).

\begin{table}
\caption{Infected countries/regions and the date of onset (dark-gray: country with local infection, light-gray: country with only imported cases) \cite{WorldHealthOrganizationRegionalOffice2006}}
\label{Date_of_Onset}
\renewcommand{\arraystretch}{1.5}
\scriptsize{
\begin{center}						
\begin{tabularx}{150mm}{Xcrrr}\toprule
\multicolumn{1}{l}{Country} &
\multicolumn{1}{c}{Date of Onset:} &
\multicolumn{1}{c}{Imported Cases} &
\multicolumn{1}{c}{Total Cases} &
\multicolumn{1}{c}{Percentage of} \\
\multicolumn{1}{c}{} &
\multicolumn{1}{c}{First Probable Case} &
\multicolumn{1}{c}{} &
\multicolumn{1}{c}{} &
\multicolumn{1}{c}{Imported Cases}
\\ \hline
\multicolumn{1}{>{\columncolor[gray]{0.6}}X}{	China	} &	16-Nov 2002	&	NA	&	5,327	&	NA	\\
\multicolumn{1}{>{\columncolor[gray]{0.6}}X}{	Hong Kong	} &	15-Feb 2003	&	NA	&	1,755	&	NA	\\
\multicolumn{1}{>{\columncolor[gray]{0.6}}X}{	Viet Nam	} &	23-Feb	&	1	&	63	&	2	\\
\multicolumn{1}{>{\columncolor[gray]{0.6}}X}{	Canada	} &	23-Feb	&	5	&	251	&	2	\\
\multicolumn{1}{>{\columncolor[gray]{0.8}}X}{	United States	} &	24-Feb	&	27	&	27	&	100	\\
\multicolumn{1}{>{\columncolor[gray]{0.6}}X}{	Taiwan	} &	25-Feb	&	21	&	346	&	6	\\
\multicolumn{1}{>{\columncolor[gray]{0.6}}X}{	Singapore	} &	25-Feb	&	8	&	238	&	3	\\
\multicolumn{1}{>{\columncolor[gray]{0.6}}X}{	Philippines	} &	25-Feb	&	7	&	14	&	50	\\
\multicolumn{1}{>{\columncolor[gray]{0.8}}X}{	Australia	} &	26-Feb	&	6	&	6	&	100	\\
\multicolumn{1}{>{\columncolor[gray]{0.8}}X}{	Ireland, Republic of	} &	27-Feb	&	1	&	1	&	100	\\
\multicolumn{1}{>{\columncolor[gray]{0.8}}X}{	United Kingdom	} &	1-Mar	&	4	&	4	&	100	\\
\multicolumn{1}{>{\columncolor[gray]{0.8}}X}{	Germany	} &	9-Mar	&	9	&	9	&	100	\\
\multicolumn{1}{>{\columncolor[gray]{0.8}}X}{	Switzerland	} &	9-Mar	&	1	&	1	&	100	\\
\multicolumn{1}{>{\columncolor[gray]{0.8}}X}{	Thailand	} &	11-Mar	&	9	&	9	&	100	\\
\multicolumn{1}{>{\columncolor[gray]{0.8}}X}{	Italy	} &	12-Mar	&	4	&	4	&	100	\\
\multicolumn{1}{>{\columncolor[gray]{0.8}}X}{	Malaysia	} &	14-Mar	&	5	&	5	&	100	\\
\multicolumn{1}{>{\columncolor[gray]{0.8}}X}{	Romania	} &	19-Mar	&	1	&	1	&	100	\\
\multicolumn{1}{>{\columncolor[gray]{0.8}}X}{	France	} &	21-Mar	&	7	&	7	&	100	\\
\multicolumn{1}{>{\columncolor[gray]{0.8}}X}{	Spain	} &	26-Mar	&	1	&	1	&	100	\\
\multicolumn{1}{>{\columncolor[gray]{0.8}}X}{	Sweden	} &	28-Mar	&	5	&	5	&	100	\\
\multicolumn{1}{>{\columncolor[gray]{0.6}}X}{	Mongolia	} &	31-Mar	&	8	&	9	&	89	\\
\multicolumn{1}{>{\columncolor[gray]{0.8}}X}{	South Africa	} &	3-Apr	&	1	&	1	&	100	\\
\multicolumn{1}{>{\columncolor[gray]{0.8}}X}{	Indonesia	} &	6-Apr	&	2	&	2	&	100	\\
\multicolumn{1}{>{\columncolor[gray]{0.8}}X}{	Kuwait	} &	9-Apr	&	1	&	1	&	100	\\
\multicolumn{1}{>{\columncolor[gray]{0.8}}X}{	New Zealand	} &	20-Apr	&	1	&	1	&	100	\\
\multicolumn{1}{>{\columncolor[gray]{0.8}}X}{	Korea, Republic of	} &	25-Apr	&	3	&	3	&	100	\\
\multicolumn{1}{>{\columncolor[gray]{0.8}}X}{	India	} &	25-Apr	&	3	&	3	&	100	\\
\multicolumn{1}{>{\columncolor[gray]{0.8}}X}{	Macao	} &	5-May	&	1	&	1	&	100	\\
\multicolumn{1}{>{\columncolor[gray]{0.8}}X}{	Russian Federation	} &	5-May	&	1	&	1	&	100	\\
\bottomrule
\end{tabularx}						
\end{center}
}
\end{table}

\begin{figure}
\centering
\includegraphics[width=12cm, angle=0]{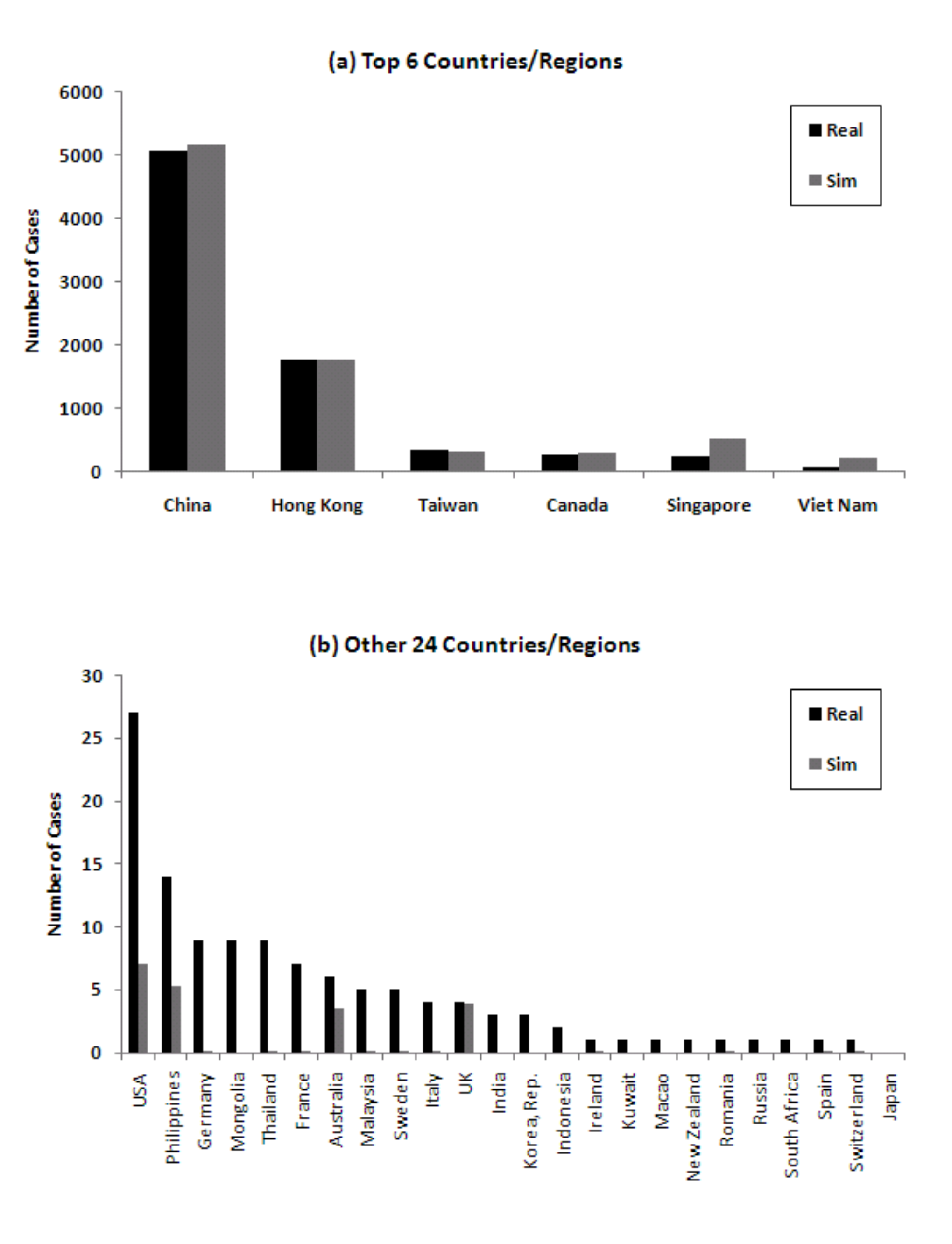}
\caption{Comparison of number of cumulative cases in 30 countries/regions considering actual time of first infection ((a) Top 6 countries/regions, (b) 24 other countries/regions)}
\label{Comp_Case_SARS_Timing}
\end{figure}

To provide some more weight to this hypothesis, we ran the model, but this time using the actual time of first infection in the country rather than travel data. Other than that, the parameters of the simulation remained the same. Figure~\ref{Comp_Case_SARS_Timing} shows that the cumulative number of cases from the model then correspond better to real data. 

\paragraph{Local Considerations}

Our model trades off adaptability to local conditions for a smaller number of parameters to fit. This can have several effects. Here we discuss two of them, and how they might affect the results. First, if we look at data from seasonal flu cases, we find that Canada typically has a large number of cases, and the United States has the largest number of influenza isolates \cite{WorldHealthOrganizationGlobalAtlas}. Both of these suggest that the local infection coefficient may be higher in Canada and the United States than other countries. Indeed, this could have been an additional factor in the surprisingly large number of Canadian cases. However, the United States was surprising, because, although it imported 27 cases, the infection did not spread locally. This may indicate that the quarantining and isolation measures employed worked effectively.

A second interesting point is that Singapore and Vietnam both report many fewer cases than predicted by the model. This may be partly explained by their lower propensity to spread infection, again as evidenced by seasonal flu data. There may also have been a significantly bigger push to hospitalize and keep patients confined, weakly evidenced by the fact that the proportion of those infected who were healthcare workers in these two countries (41\% and 57\% in Singapore and Vietnam respectively) was much higher than other countries (21\%).

%% file: discussion.tex
We have discussed a hybrid network and local model for the spread of pandemics, and applied it to the case of SARS. When parameters are calibrated to real data on populations, densities, and traffic, we show that the model reproduces many of the key dynamics of the spread of SARS in 2002 and 2003, while remaining parsimonious, and therefore useful for understanding the root causes of why pandemics spread in the way they do. Both the successes and the failures of the simple model provide insights into pandemic spread. For example, it is clear that it is important to model international traffic to understand the pathways of spread. At the same time, for any particular pandemic, individual idiosyncrasies can come into play. For example, the importance of super-spreaders in SARS is reflected in the fact that the time of first infection in a country has a big role in how many people get infected. The other major takeaway from this work is that the level of granularity in the network structure of the model has a significant impact. For example, treating China as one large entity leads to poorer prediction, but at the same time specializing all the way down to cities would end up requiring too much data to accurately calibrate the model, and would probably not provide significantly better prediction.